# MARVEL, a four-telescope array for high-precision radial-velocity monitoring

G. Raskin*[a,b], C. Schwab[c], B. Vandenbussche[a], J. De Ridder[a], C. Lanthermann[a], J. Pérez Padilla[a,b], A. Tkachenko[a], H. Sana[a], P. Royer[a], S. Prins[a,b], L. Decin[a], D. Defrère[a], J. Pember[b], D. Atkinson[d], A. Glasse[d], D. Pollacco[e], G. Tinetti[f], M. Güdel[g], J. Stürmer[h], I. Ribas[i,j], A. Brandeker[k], L. Buchhave[l], S. Halverson[m], G. Avila[n], J. Morren[o], H. Van Winckel[a]

[a] Institute of Astronomy, KU Leuven, 3001 Leuven, Belgium
[b] Mercator Telescope, Roque De Los Muchachos Observatory, La Palma, Spain
[c] Department of Physics and Astronomy, Macquarie University, NSW 2109, Australia
[d] UKATC, Blackford Hill, Edinburgh EH9 3HJ, United Kingdom
[e] Department of Physics, University of Warwick, Coventry CV4 7AL, UK
[f] Centre for Space Exoplanet Data (CSED), University College London, London WC1W 6BT, UK
[g] Department of Astrophysics, University of Vienna, 1180 Vienna, Austria
[h] ZAH Landessternwarte, Heidelberg, Germany
[i] Institut de Ciències de l'Espai (ICE, CSIC), Campus UAB, E-08193 Bellaterra, Barcelona, Spain
[j] Institut d'Estudis Espacials de Catalunya (IEEC), E-08034 Barcelona, Spain
[k] Department of Astronomy, AlbaNova University Center, Stockholm University, Sweden
[l] DTU Space, National Space Institute, Technical University of Denmark, Kgs. Lyngby, Denmark
[m] Jet Propulsion Laboratory, California Institute of Technology, Pasadena CA USA
[n] optical Lab, Baader Planetarium GmbH, 82291 Mammendorf, Germany
[o] Department of Physics and Astronomy, KU Leuven, 3001 Leuven, Belgium

## ABSTRACT

Since the first discovery of a planet outside of our Solar System in 1995, exoplanet research has shifted from detecting to characterizing worlds around other stars. The TESS (NASA, launched 2019) and PLATO mission (ESA, planned launch 2026) will find and constrain the size of thousands of exoplanets around bright stars all over the sky. Radial velocity measurements are needed to characterize the orbit and mass, and complete the picture of densities and composition of the exoplanet systems found. The Ariel mission (ESA, planned launch 2028) will characterize exoplanet atmospheres with infrared spectroscopy. Characterization of stellar activity using optical spectroscopy from the ground is key to retrieve the spectral footprint of the planetary atmosphere in Ariel's spectra. To enable the scientific harvest of the TESS, PLATO and Ariel space missions, we plan to install MARVEL as an extension of the existing Mercator Telescope at the Roque De Los Muchachos Observatory on La Palma (SPAIN). MARVEL consists of an array of four 80 cm telescopes linked through optical fibers to a single high-resolution echelle spectrograph, optimized for extreme-precision radial velocity measurements. It can observe the radial velocities of four different stars simultaneously or, alternatively, combine the flux from four telescopes pointing to a single faint target in one spectrum. MARVEL is constructed by a KU Leuven (Belgium) led collaboration, with contributions from the UK, Austria, Australia, Sweden, Denmark and Spain. In this paper, we present the MARVEL instrument with special focus on the optical design and expected performance of the spectrograph, and report on the status of the project.

---

* gert.raskin@kuleuven.be

## 1. INTRODUCTION

Since the first discovery of an exoplanet around a Sun-like star [1], the research field of exoplanets has grown rapidly and is now more active than ever. In particular, several ambitious ESA and NASA space missions are shaping the future of exoplanet research: CHEOPS will measure the size of exoplanets with known mass [2]. TESS [3] and PLATO [4] will discover thousands of new transiting exoplanets, including smaller planets than those currently possible to detect, with the aim to identify several Earth analogues. Ariel [5] will observe the atmospheric spectra of hundreds of exoplanets. The full scientific exploitation of these space data will demand a large amount of supporting radial-velocity measurements from the ground, for which dedicated infrastructure is pivotal. MARVEL intends to precisely fill that niche.

The design of MARVEL is specifically targeted towards the radial velocity (RV) follow up of the PLATO mission of the European Space Agency (ESA) that will be launched in 2026. PLATO will deliver long-term (> 2 years per field) photometric monitoring of 2232 deg$^2$ sky fields to find and study exoplanet systems. The mission will observe large fields in the sky in order to access a large sample of bright stars (85000 stars with $m_V$ < 11), that allow spectroscopic follow-up from the ground with relatively small telescopes. From the PLATO transit measurements, the relative size of the exoplanet with respect to the host star can be established. In order to obtain densities, mass determination via RV measurements is needed. The expected yield of planet host candidates that require multi-epoch (typically 30 – 100 [6]) RV follow up to determine orbit and mass, is about 4000 stars. The required precision of these measurements is at the 1 m/s level. Between 2 and 120 of these systems are expected to contain Earth-like planets in the habitable zone, for which even more accurate sub-m/s RV follow-up is planned on large telescopes (e.g. ESPRESSO, HARPS). With MARVEL we aim at providing a single facility to deliver a major part of the PLATO P1 and P2 sample candidates follow-up, able to perform about 20 000 1-m/s measurements per year. The PLATO P1 sample includes dwarf and subdwarf stars (F5-K7) with $m_V$ < 11. The P2 sample includes dwarf and subdwarf stars (F5-K7) with $m_V$ < 8.2. Combining multi-epoch high-precision RV measurements from a single facility will allow us to avoid instrument biases when combining measurements from a heterogeneous collection of telescopes. While 1 m/s follow-up capability is currently achievable on 4-meter class telescopes, the availability of time on these facilities is limited.

The ESA Ariel mission was selected as part of ESA's Cosmic Vision program. It will study what exoplanets are made of, how they formed and how they evolve, via visible and infrared transit and eclipse photometry and spectroscopy. The Ariel target list covers a diverse sample of about 1000 exoplanets. Candidate targets need to be characterized for target list optimization, and high-resolution spectroscopy will be indispensable for all these candidate targets, to measure the host star's fundamental parameters, its chromospheric activity, and mass and orbit of the planet, if not known. A major source of measurement uncertainty for Ariel is stellar activity, especially star spots. The characterization of the host star's activity during the Ariel measurements is a crucial issue in the analysis of the transit spectra. Precision RV monitoring of the host stars hold great promise in combination with photometric monitoring [7].

Another promising application on a shorter term for MARVEL consists of the RV follow-up of TESS (Transiting Exoplanet Survey Satellite), a mission that is focusing on the nearest and brightest exoplanet hosts, and on exoplanets with short orbital periods (1 month, versus up to 1 year for PLATO). Many of the brightest TESS targets are amenable to precision RV follow-up using smaller-scale telescopes, making MARVEL a powerful tool for measuring precise densities of small TESS planets.

MARVEL will become an extension of the existing Mercator Telescope at the Roque De Los Muchachos Observatory on La Palma (Spain). It is developed and constructed by an international consortium, led by the Institute of Astronomy of KU Leuven (Belgium) and further consisting of the University of Warwick (UK), the UK Astronomy Technology Centre, the University of Vienna (Austria), Macquarie University (Australia), Landessternwarte Heidelberg (Germany), AlbaNova University Center (Sweden), Institute of Space Sciences (IEEC-CSIC, Spain), and DTU Space (Denmark). Currently, we are working towards a preliminary design review, after which we envisage a fast track to commissioning in 2023.

## 2. INSTRUMENT CONCEPT

MARVEL is dedicated to high-precision radial-velocity spectroscopy using an array of four commercially available, robotic telescopes feeding one spectrograph, a concept already employed by the MINERVA array [7], but with higher spectral resolution, higher throughput and much larger wavelength coverage. It will enable simultaneous RV measurements of multiple bright targets distributed over the sky. One telescope will be allocated to each of the four targets observed simultaneously. This way, each telescope can independently access the entire observable sky, hence providing maximum flexibility in the scheduling of observations. The four telescopes can also point to the same target, combining the observed fluxes and henceforth allowing fainter targets to come into operational range. One of the key aspects of the MARVEL concept is that it allows us to use small and hence relatively cheap telescopes, feeding a single highly efficient, RV-optimized spectrograph. Not only is the cost of four 80-cm telescopes substantially lower than a single 1.6-m telescope with the same light collecting power, smaller telescopes also allow us to build a more efficient and/or less expensive spectrograph without having to resort to pupil or image slicing. Indeed, the dimensions of the optical elements of a seeing-limited astronomical spectrograph scale with the diameter of the telescope it is attached to. This is a consequence of the optical principle of conservation of étendue. The limited size of the individual MARVEL telescopes is therefore an advantage as it allows us to design a compact instrument with high throughput, that can actually be competitive with comparable instruments on 4-meter class telescopes at only a fraction of their cost. Furthermore, operational overheads (e.g. target acquisition time) are in general longer on larger, more complex facilities. Stellar pulsations also affect the accuracy of the RV measurements. This effect can be mitigated by averaging out multiple pulsation periods over a sufficiently long measurement (typically on the order of 15 minutes). However, this severely limits the potential reduction of exposure time when using much larger telescopes and targeting rather bright stars, as is the case for PLATO and Ariel. In other words, what is lost in light collecting power of the small telescopes is partly compensated by the gain in spectrograph throughput and observing efficiency, as well as by the multi-object capability of the telescope array, provided that the design is also optimized for high throughput.

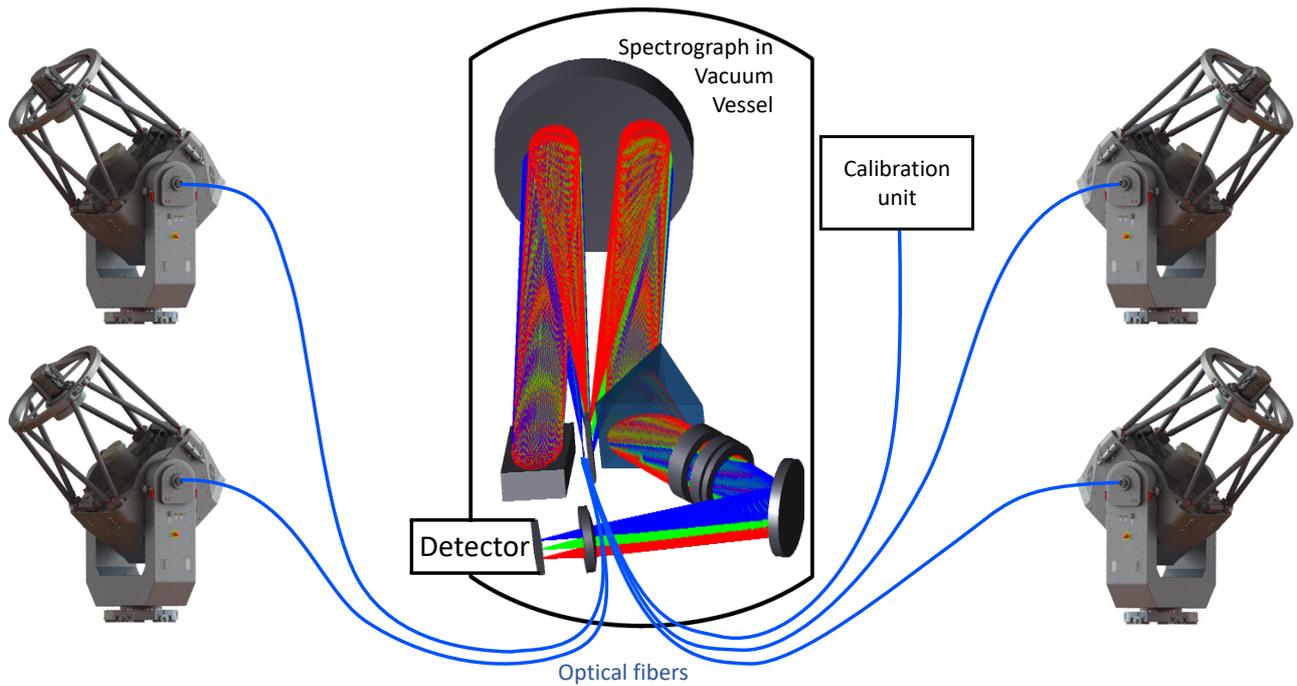

Figure 1. Conceptual layout of MARVEL.

Figure 1 shows the conceptual layout of the MARVEL facility. It consists of four commercially-off-the-shelf (COTS) telescopes with a diameter of 80 cm. Each telescope will be equipped with a fiber adapter front end and coupled through an optical fiber to a common high-resolution echelle spectrograph. This instrument will have a spectral resolution $R = \lambda/\Delta\lambda$ of at least 90 000 and cover a very large wavelength range in a single exposure, spanning at least 390 nm (to include the calcium H and K lines) to 920 nm. The availability of very large CCD detectors allows us to record the spectra of four stars simultaneously with a single fiber-fed instrument as long as the targets have a similar brightness, such that a common exposure time can be used. The four stellar spectra are interlaced with a fifth spectrum from a wavelength calibration source that is used to track the instrumental drift during each exposure to improve the RV measurement.

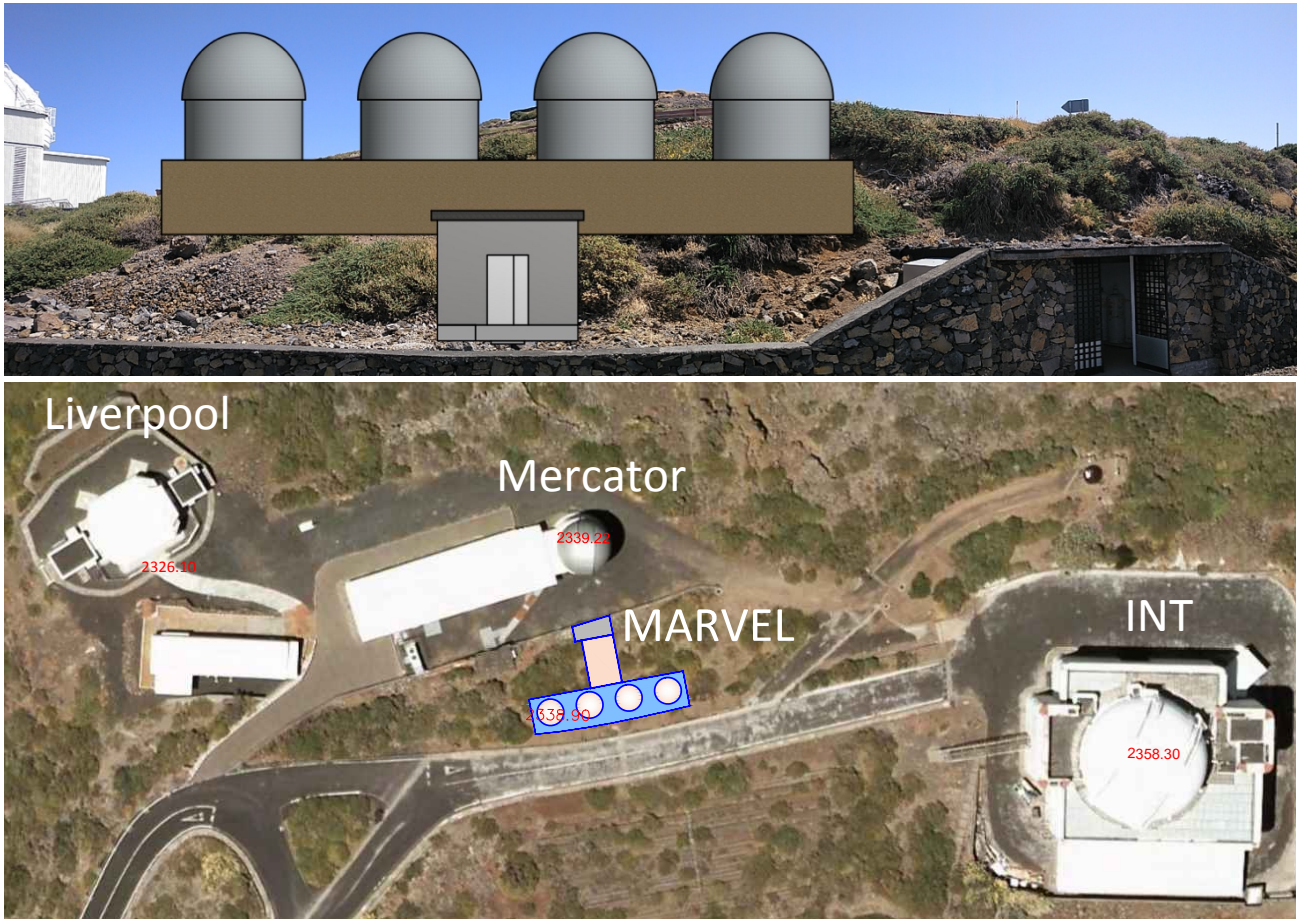

Figure 2. Rendering of MARVEL as seen from the Mercator Telescope; Bottom: aerial view of the MARVEL location and of the neighboring telescopes (altitude of the facilities is indicated in red).

In a parallel project, part of the MARVEL consortium is well advanced in the development of a wavelength calibration system based on the transmission spectrum from a Fabry-Perot etalon, illuminated with light [8]. This spectrum consists of a dense comb of spectral peaks, regularly spaced in frequency. To ensure absolute stability of the calibration spectrum, one of the etalon peaks is referenced to a rubidium hyperfine transition line through laser spectroscopy. This system is currently under commissioning at the Nordic Optical Telescope on La Palma and will soon be deployed at the HERMES spectrograph on the Mercator Telescope too. The Fabry-Perot etalon spectrum will be the core of the MARVEL wavelength calibration. Furthermore, the emission line spectrum from a Thorium-Argon hollow cathode lamp will be available as a back-up solution or for cross-referencing the etalon spectrum.

Both ambient temperature and pressure affect the stability of the instrument. Temperature variations cause thermal expansion of the optics as well as fluctuations in the refractive index of glass and air. The latter also depends on air pressure. Both effects result in a spectral shift that mimic a radial-velocity signal. To suppress the effects of ambient pressure variations, the MARVEL spectrograph will be installed in a large vacuum vessel with active temperature control of the optical bench inside. Furthermore, this vacuum vessel and sensitive control electronics will be housed in a well-insulated room with very precise temperature control. We envisage a thermal stability of the spectrograph optical bench at the milli-Kelvin level.

MARVEL operations will be fully robotic, but for operational efficiency and to allow sharing infrastructure, MARVEL will be installed close to the Mercator Telescope at the Roque De Los Muchachos Observatory. The exact location is illustrated in Figure 2. Each 80-cm telescope will be housed in a 4-m diameter dome. The spectrograph will be installed in a small well-insulated building next to the telescope platform.

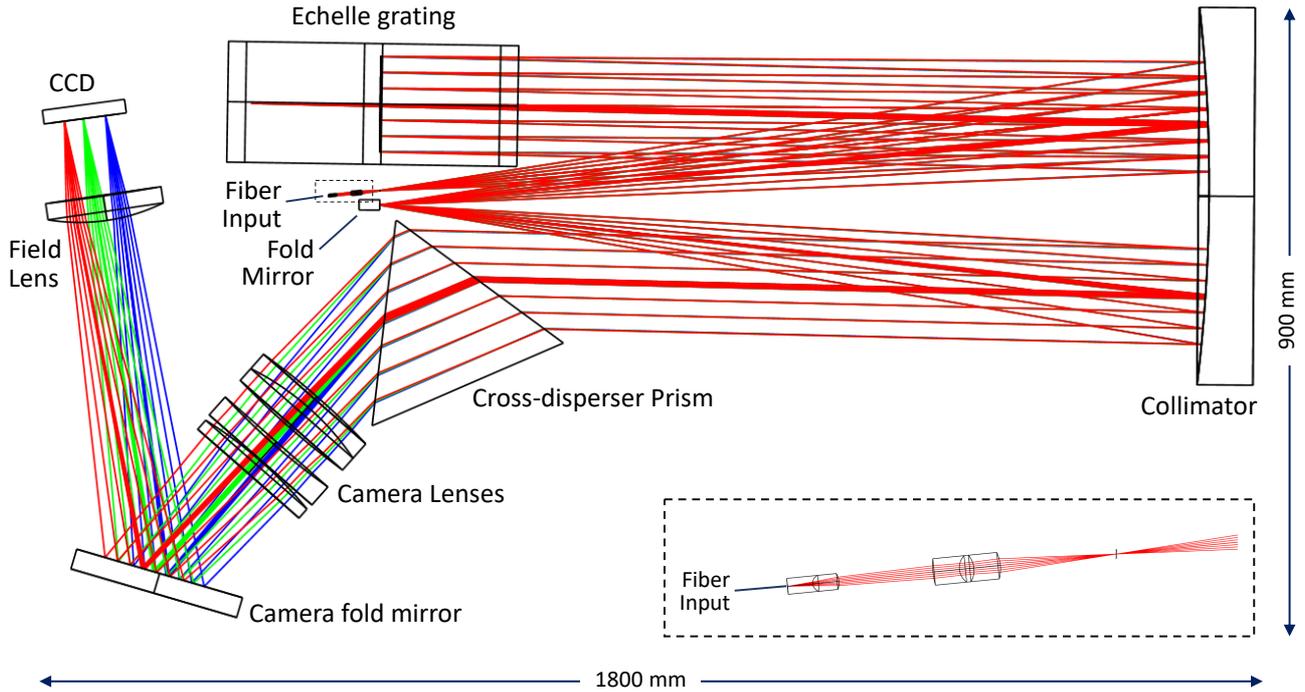

Figure 3. Optical layout and ray-trace of the MARVEL spectrograph. The dashed box in the lower right corner shows in detail the layout of the focal ratio adaptation optics at the entrance of the spectrograph; five fibers are glued on the entrance surface of the first lens.

## 3. SPECTROGRAPH DESIGN

The optical design of the MARVEL spectrograph, shown in Figure 3, is strongly based on NEID [9], the extreme precision Doppler spectrograph for the WIYN telescope. It follows a traditional white-pupil layout based on a monolithic parabolic mirror used in triple pass. The grating tilt ($\gamma = 0.5°$ perpendicular to the plane of dispersion) to separate the dispersed from the incident beam is kept as small as possible, narrowing the interspace between the fiber input and the fold mirror to a strict minimum. This limits the size of the collimator mirror and relaxes its focal ratio as well as helps keeping the overall dimensions of the spectrograph and the surrounding vacuum vessel small.

The spectrograph is fed by five octagonal-core fibers (4 science fibers, 1 wavelength reference fiber) with a width of 40 µm. At the spectrograph entrance, the focal ratio of the beam exiting the fibers is converted from $F/4$ to $F/8$ by a set of focal ratio adaption optics, consisting of one doublet and one triplet lens. To increase throughput, the 5 fibers are aligned and cemented on the flat entrance surface of the first lens. Outside the vacuum vessel, the four octagonal science fibers are butt-to-butt coupled and accurately aligned to four circular fibers that run to the four telescopes. Combining circular and octagonal fibers breaks the circular symmetry of the former and increases the overall scrambling performance of the fiber link without the throughput penalty of adding an optical double scrambler. This technique was already used successfully in CARMENES [10][11]. It results in improved image and pupil stability in the spectrograph with respect to a single fiber, and hence, better RV accuracy. The 40-µm fiber diameter in the focal plane of each telescope corresponds to a comfortably large sky aperture of 2.3 arcsec without having to resort to image or pupil slicing. Again, this contributes to the overall MARVEL performance.

MARVEL's main disperser is a Richardson Grating Labs (RGL) echelle grating with a 69° blaze angle (R2.7) and a groove spacing of 52.67 lines/mm. It is a Zerodur replica from RGL's master ruling MR138 that was already used successfully for the HERMES spectrograph [12]. The R2.7 echelle is a compromise between the smaller beam diameter of an R4 and the smaller angular dispersion of an R2 echelle. The latter proportionally reduces the cross-dispersion requirements. This is essential to obtain sufficient inter-order separation for fitting five separate spectra. With a ruling width of 408 mm, a beam diameter of 146 mm exactly fills the projected grating aperture. However, we decided to use a 160 x 408 mm grating to accommodate a 160-mm diameter beam. This allows us to recover part of the flux that otherwise might be lost due to focal ratio degradation (up to 16%) while accepting some limited overfilling of the grating (~3%). For the same reason, the cross-disperser and camera optics are slightly oversized. This dispersion geometry, in combination with a 40-μm core fiber, provides MARVEL with a spectral resolution of at least $R = 90\,000$.

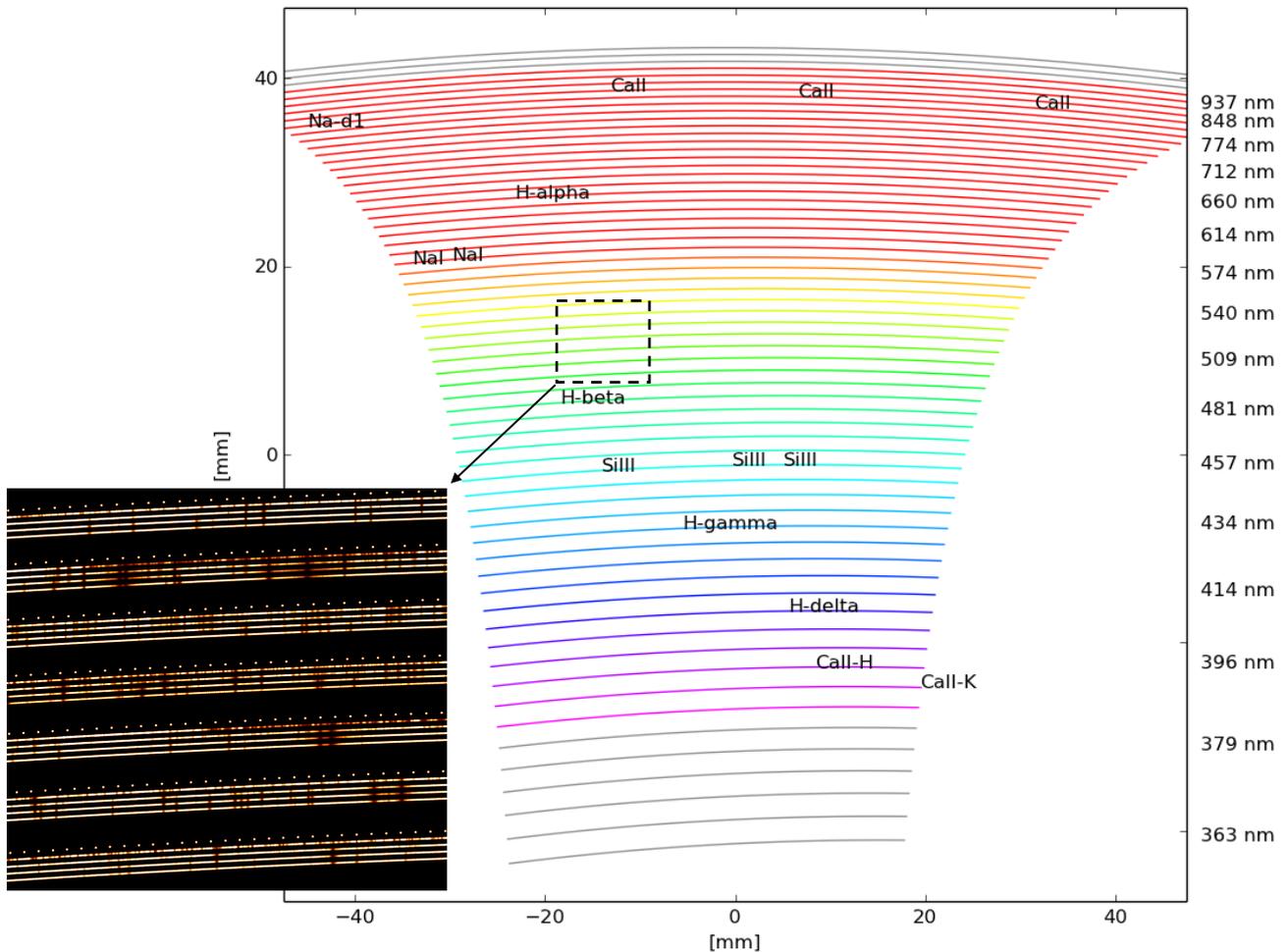

Figure 4. Layout of the 2D spectrum on the STA1600 detector. The reddest orders are slightly truncated short of the free-spectral-range. The location of some important spectral lines is indicated in the echellogram, as well as the center wavelength of each 4$^{th}$ order at the right. Image quality and throughput are optimized for the colored orders and can degrade for the grey orders. The inset at the lower left shows part of 7 spectral orders from an Echelle++ simulation [13] of 4 stellar spectra and of 1 Fabry-Perot wavelength reference spectrum.

As mentioned in the previous paragraph, sufficient cross dispersion is a critical issue of the MARVEL spectrograph. Because of the very wide spectral range – spanning more than 1.25 octaves – a diffraction grating is excluded as cross disperser, and we have to rely on a prism. In terms of throughput, especially at the edges of the spectral range where grating efficiency falls off steeply, a large prism is the preferred solution as well. Moreover, a prism produces hardly any cross-order scattered light and offers more evenly spaced orders, as compared to a grating. MARVEL uses one very large prism (60° apex angle, 54.6° angle of incidence, entrance/exit surface dimensions: 200 x 300 mm) as cross disperser, made out of F2HT (Schott) or alternatively PBM2Y (Ohara) glass. These are the highest-dispersion glasses commercially available with still reasonably good transmission at 390 nm (>99.5%/cm). The cross disperser provides an order separation ranging from 240 pixels at 390 nm to 82 pixels at 900 nm, sufficient to accommodate five interlaced spectra.

The MARVEL detector is a thinned back-illuminated STA1600 CCD with 10.3k x 10.3k 9-μm pixels (95 x 95 mm image area). The camera optics are designed to image five interlaced full-range spectra on this detector with just a few minor gaps at the edges of the reddest spectral orders (Figure 4). Each resolution element is sampled by 5 pixels on average. This allows us to maintain sufficient sampling when switching to 2 x 2 binning to lower the magnitude limit in case of observing faint targets, or for low SNR exposures. However, most observations will be carried out at high SNR, hence the increased total read noise due to the 5-pixel sampling will not affect performance. On the contrary, the generous sampling will improve spectral fidelity and reduce the effect of systematic errors like e.g., detector inter and intra-pixel inhomogeneities.

The fully refractive camera ($F$ = 750 mm, $F/4.7$) comprises two lens groups: a front element, consisting of one cemented doublet and two singlets, and one field lens at a comfortable back focal distance from the detector. In between, a fold mirror is added to reduce the vacuum tank dimensions. We abandoned the use of a cylindrical field flattener close to the detector for correcting the field curvature introduced by the white-pupil mounting. In exchange, the field curvature is effectively canceled by a small decenter and tilt in cross-order direction of the field lens.

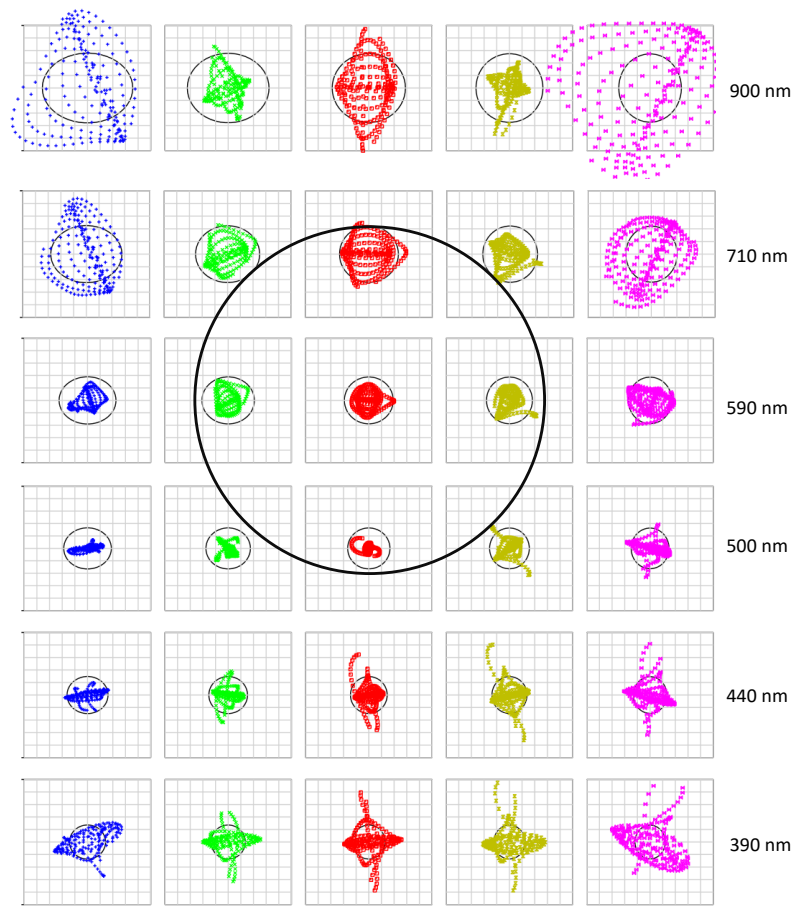

Figure 5. Spot diagrams for the central fiber position in 6 spectral orders, showing 5 wavelengths per order spread evenly over the free spectral range. The square boxes measure 18 μm, corresponding with 2 x 2 detector pixels. The ellipses inside each box indicate the Airy disk diameter, their change in elongation illustrating the anamorphism of the dispersion optics. The large circle in the center corresponds with a 40-μm fiber diameter projected on the detector (same scale as the spot diagram boxes).

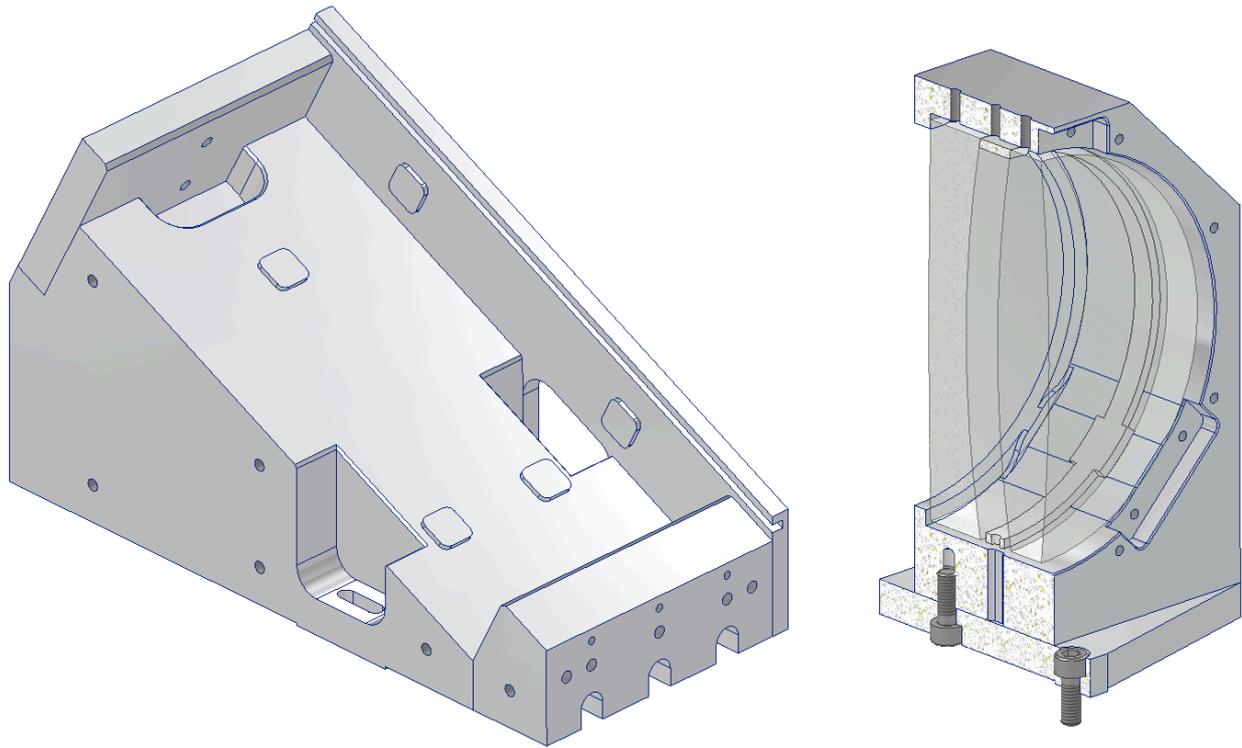

Figure 6. Examples of the optomechanical mounts for the echelle grating (left) and the camera doublet (right).

This camera, although consisting of only five all-spherical lenses, and only relying on glasses with good UV transmission, offers exceptional image quality over the complete spectral range, as can be appreciated from the spot diagrams in Figure 5. For all but the most extreme wavelengths, the diffraction limit is approached. Apart from improving spectral resolution and producing well-defined and clean cross-order profiles, excellent image quality is favorable for the RV accuracy too. Indeed, in case of imperfect optics, any variation in the illumination of the pupil leads to a different weighting of the aberrated point spread function, and hence to a wavelength-dependent shift of the spectral lines, which causes spurious RV errors.

The optomechanical design of MARVEL builds on the heritage of NEID, HERMES and smaller spectrographs developed at Macquarie University[14], using aluminum mounts – monolithic where possible – and fully kinematic mounting for utmost mechanical stability. The high thermal conductivity of the solid aluminum parts aids to achieve the lowest possible thermal gradients. Two examples of this type of mounts are shown in Figure 6.

## 4. PERFORMANCE

Given the small collecting area of the MARVEL telescopes, outmost attention is paid to the optical efficiency of the instrument. The number of optical surfaces is kept at a minimum, focal ratio degradation (FRD) losses in the fiber link will be small thanks to oversizing the spectrograph beam, and high-performance mirror and anti-reflection coatings will be used. The left panel of Figure 7 shows the ambitious but realistic throughput targets of the main MARVEL components and of the complete instrument. The spectrograph efficiency values are calculated at the blaze peaks of the echelle grating. For the coupling efficiency of the star image to the fiber aperture, we assumed an atmospheric seeing of 1", well-above the median seeing of 0.8" at the Roque De Los Muchachos Observatory. Atmospheric extinction is not included in the total throughput. Over the wavelength range from 450 to 600 nm, the total throughput peaks at 25%.

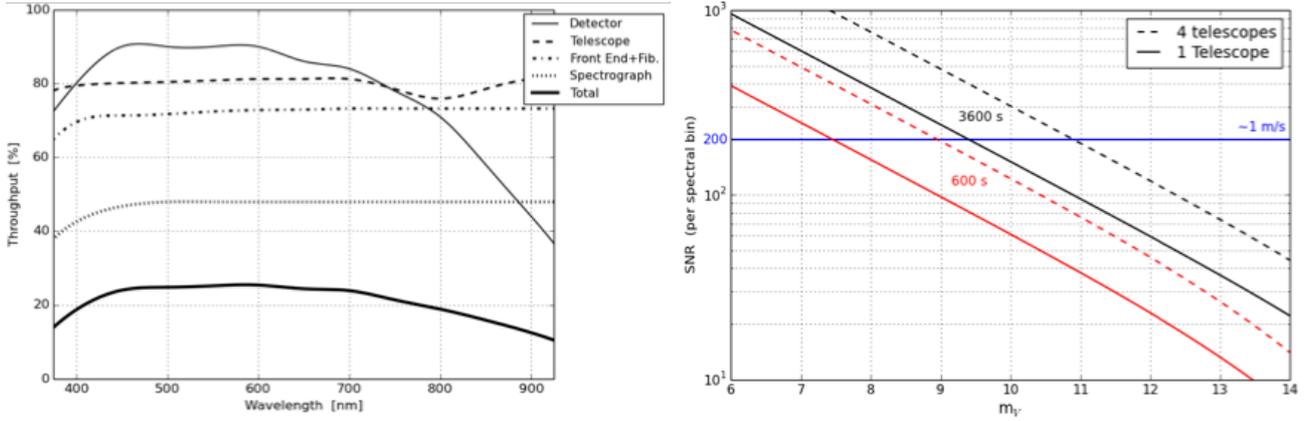

Figure 7. Left: Throughput of the main components and of the total MARVEL instrument. Right: Estimated peak SNR per resolution bin at 550 nm for a range of magnitudes, for a single 80-cm telescope and for 4 telescopes pointing to the same target; exposure times are 10 minutes (red) and one hour (black).

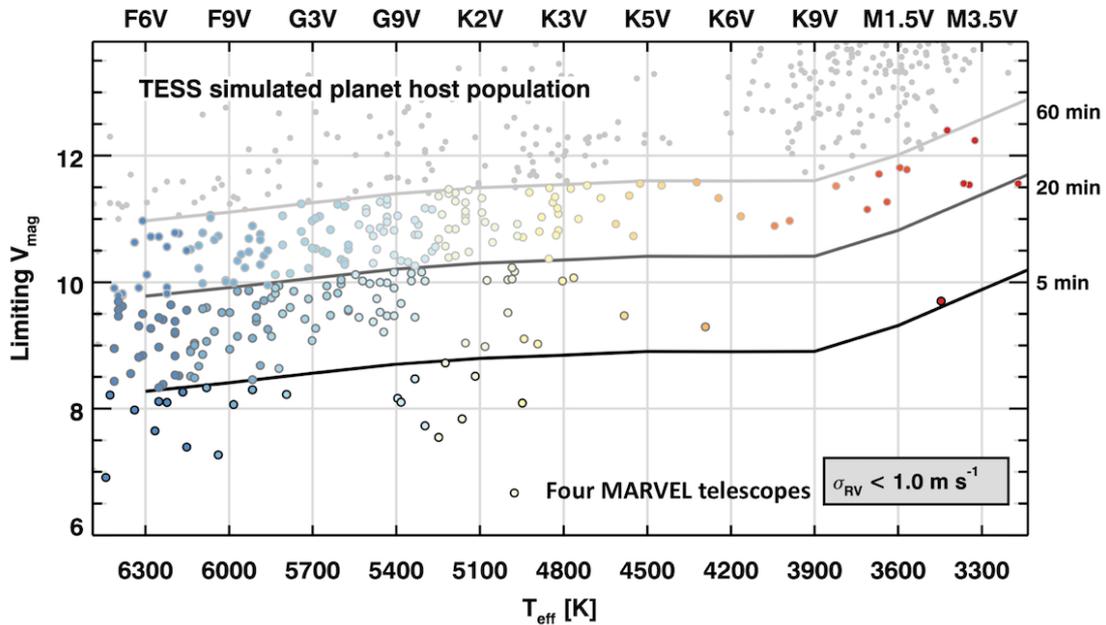

Figure 8. Estimated MARVEL V magnitude limits for achieving an RV uncertainty below 1 m/s with exposure times of 5 and 20 minutes and 1 hour, for four telescopes pointing to the same target (when using a single telescope, the magnitude limits worsen by 1.5 mag). The expected TESS planet-hosting stars are overplotted.

Based on these estimated throughput values, we calculated the signal-to-noise ratio (SNR) that we expect to achieve with MARVEL at a spectral resolution of R = 90 000, using one or four 80-cm telescopes (Figure 7, right panel). These SNR values are calculated at a wavelength of 550 nm for a range of magnitudes. From a star of $m_V$ = 8.9, we can obtain a SNR of 200 per resolution element in a 1-hour exposure with a single telescope. In case we point all four telescopes to the same target and integrate their fluxes, this magnitude limit increases to $m_V$ = 10.9. These magnitude values fit nicely within the brightness range of the PLATO and Ariel targets. An SNR of around 200 is sufficient to suppress photon noise to the level required for attaining the 1 m/s RV precision limit on a host star of spectral type F. The required SNR decreases for cooler stars. In Figure 8, we show the magnitude limit for which the photon noise drops to the 1-m/s level as a function of spectral type and exposure time, calculated according to the method described in [15]. In a conservative approach, we based this calculation on throughput values that are only 75% of the throughput design goals shown in Figure 7. In the simulation, we assumed that all targets are slowly rotating stars ($v \sin i = 2$ km/s). To estimate more realistic RV uncertainties, we excluded regions of the spectrum within ± 15 km/s of telluric features that have a 1% depth or greater at a spectral resolution of 100 000. The expected TESS planet-hosting stars from [16] are overplotted in Figure 8, showing that MARVEL will allow us to follow up a large sample of interesting targets with very high precision. For the faintest targets, it could be possible to accept a larger RV uncertainty of e.g. 2.5 m/s, thereby increasing the $m_V$ limit by one full magnitude. The plots of Figure 8 illustrate that MARVEL has the potential to become competitive with state-of-the-art RV instruments on 4-meter class telescopes, in particular considering the amount of time available at the facility.

## 5. CONCLUSION

In this contribution, we presented the instrument concept and optical design of MARVEL, a novel high-precision RV facility that combines high throughput with excellent RV accuracy. The combination of four small-aperture telescopes with a single high-resolution spectrograph facilitates designing and building a highly competitive and cost-efficient instrument. MARVEL is set to play a key role in enabling the scientific legacy of PLATO and Ariel, two upcoming ESA M-class missions. MARVEL will allow the characterization of the masses of a large fraction of the exoplanets found by PLATO. Furthermore, it is ideally suited to monitor the activity of the host stars of the Ariel targets, crucial for the success of this mission.

## ACKNOWLEDGMENTS

The MARVEL project and team acknowledge support from the Fund for Scientific Research of Flanders (FWO) under grant I011020N (Large infrastructure program) and a grant from the University of Vienna (investment project for MARVEL).